\documentclass[a4paper,10pt]{article}
\usepackage[utf8x]{inputenc}
\usepackage{graphics}
\usepackage{graphicx}

\title{Performance simulation of a MRPC-based PET Imaging System}
\author{A. Banerjee, S. Chattopadhyay}

\begin{document}

\maketitle

\begin{abstract}

The low cost and high resolution gas-based Multi-gap Resistive Plate Chamber (MRPC) opens a new possibility to find an efficient alternative 
detector for Time of Flight (TOF) based Positron Emission Tomography, where the sensitivity of the system depends largely on the time resolution 
of the detector. Suitable converters can be used to increase the efficiency of detection of photons from annihilation. In this work, we perform a detailed GEANT4 simulation 
to optimize the converter thickness thereby improving the efficiency of photon conversion. Also we have developed a Monte Carlo based 
simulation of MRPC response thereby obtaining the intrinsic time resolution of the detector, making it possible to simulate the final response of MRPC-based 
systems for PET imaging. The result of the cosmic ray test of a four-gap Bakelite-based MRPC operating in streamer mode is discussed.

\end{abstract}

\section{Introduction}
\label{}
Under Nuclear Medicine, Positron Emission Tomography (PET) is an important imaging technique in which a radio-pharmaceutical having 
positron emitter is injected into the object of study. This radioactive element can be used to collect the morphological information
from the object. Inside the object, these positrons annihilate with the available electrons to produce two nearly back-to-back photons each 
having energy of 511 keV. The simultaneous detection of these photons with efficient detectors can lead to the identification of annihilation point.
As these positron emitters are injected via some physiological substances, mapping of the density of the positron sources can give 
the information about the activity inside the object of interest.

Although PET gives various important information which the other imaging techniques are not able to provide, limitations due to a short field
of view (FOV)($\sim$16 cm in standard PET scanners), long dead time of the electronics among others affect the sensitivity and the spatial resolution
of the currently available PET scanners. Moreover, to obtain an image, in a reasonable examination time a high radiation dose is required to 
deliver within the object to collect sufficient amount of data.

In a Time-of-Flight (TOF) PET scanner, the arrival times of each coincident pair of photons are more precisely known, and the 
(extremely small) difference between them is used to localize the annihilation event along the line between the two detected photons. 
This additional TOF information helps to improve the image quality by reducing the length over which the coincident events are back-projected.
Instead of being back-projected over the entire distance through the body, the events are back-projected over a smaller distance,
determined by the timing resolution of the scanner. For example, for a coincidence timing resolution of 600 ps the positional 
uncertainty is 9 cm Full Width at Half Maximum along the line pair. The average diameter of a typical patient (torso) is 25 - 30 cm, 
or approximately three times the TOF positional uncertainty. The ratio of patient diameter (D) to the positional uncertainty ($\Delta$x) has 
been reported to be representative of the noise reduction, or the sensitivity gain, with TOF. This benefit in TOF increases as the 
PET system timing resolution improves.

The detectors that are generally used worldwide for PET imaging are the scintillator-based detectors, e.g., Bismuth germanate (BGO) or Lutetium 
orthosilicate (LSO) , as they have excellent light output and very good energy resolution. But because of its high cost and the above-said 
limitations, several attempts are being made to find an alternative. Multi-gap Resistive Plate Chambers (MRPCs) are considered to be a good 
alternative to the scintillator-based PET imaging systems because of their excellent time resolution ($\sim$ 50 ps) and
 position ($\sim$ 1 mm) resolution in addition to the relatively lower cost of fabrication. In MRPC-based system, the location of the origin of 
photons can be obtained in 3-dimensions. The detector cell provides the position in transverse plane (x-y plane) and the time of arrival on the 
detector gives the distance in z-direction. If the time resolution of the MRPC is known then an estimation of the spread for the position of origin 
of photon source can be obtained by the formula \cite{1}:

\begin{equation}
  \Delta L [mm] \approx \sigma_{t}[ps]/2
\end{equation}

where $\Delta L$ is the FWHM in the position accuracy and $\sigma_{t}$ is the rms time accuracy per photon. Reconstruction of events  
however improves the resolution therby producing the desired image granularity (few mm).

Several attempts have been made in building MRPC-based prototype PET imaging systems using varying number of gaps. However, like many other 
gas-based detector systems, the efficiency of conversion of 511 keV photons is very low. MRPC, due to its layered structure, has a natural 
configuration where proper conversion materials can be placed to obtain better efficiency. However, very thick converter will likely to absorb 
the produced electrons before they can reach the gas volume, whereas for very thin one it is very difficult to have a good conversion efficiency. 
Efforts have been made to optimize the converter-MRPC combination for obtaining best possible efficiency of single 511 keV photon detection. 
A maximum value up to $\sim$ 20-25\% has been reported, as mentioned  in Ref \cite{1,4}.

In this work, we have performed a detailed simulation to optimize the detection efficiency of a pair of 511 keV photons likely
to be generated from positron annihilation. The work includes the response of the passage of the photons inside the material 
by GEANT4 and a Monte-Carlo (MC) based simulation package for obtaining the MRPC signals giving their timing properties due to produced 
electrons. The simulation is done in two stages. First, GEANT4 has been used to simulate the conversion of the incident photons. The electrons
obtained by photon conversion are considered to be the particles ionizing the gas. The processes for ionization to signal generation have been 
implemented in another MC code.

We organize the article as follows. In Sec. 2, we discuss the simulation procedure, in Sec. 3 we give the results obtained from the 
simulation and in Sec. 4 we present the results of the performance of a four gap Bakelite-base MRPC, operated in streamer mode.

\section{Simulation Procedure}
\label{}

\subsection{Conversion of 511 keV photons}
\label{}

GEANT4 (version 4.9.4) has been used to simulate the response of photons through the detector thereby generating electrons which 
reach the gas volume. GEANT4 \cite{6,7} is the efficient toolkit where different aspects of simulation process, such as the geometry 
of the system, the materials involved, the generation of secondary particles, the tracking of particles through materials, 
the physics processes governing particle interactions, the response of sensitive detector components,etc. can be implemented. For gamma
ray conversion, the data file G4EMLOW version 6.19, containing cross-sections for low energy electromagnetic processes, was used.

\begin{figure}[h!]
\begin{center}
\includegraphics[scale=0.30]{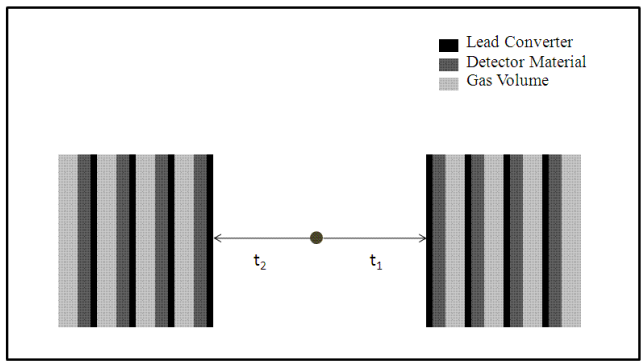}\\
\caption{\label{fig:epsart}Schematic representation of the detector configuration} 
\label{fig:1}
\end{center}
\end{figure}

The photon-pairs are considered to be back to back lying on the same plane. For attaining good conversion efficiency, we have used a set of 
Lead-MRPC combinations as shown schematically in Fig. \ref{fig:1}. Two such combinations are placed facing each other considering the source 
of photon-pair to be in between as shown in Fig. \ref{fig:1}. In this work, as an input of GEANT4, we have used different materials including 
lead as the converter, glass as detector material and the mixture C$_{2}$F$_{4}$H$_{2}$/i-C$_{4}$H$_{10}$/SF$_{6}$(85/5/10) as the gas volume 
in the gas gap. We have studied the variation of the conversion efficiency by varying the converter thickness.

Two photons with momenta of same magnitude (511 keV) but opposite directions are used as the input for every event. Any particle 
(photon or electrons) backscattered from one detector to the other are not considered for further study. Photons get converted mainly 
in the lead volume mostly by the Compton and the photoelectric processes \cite{1,2} . A number of low energy electrons get absorbed inside
the lead volume. Only the electrons reaching any one of the gas volume are considered for signal generation. Increasing the number of Pb-MRPC 
combinations improves the conversion probability, which subsequently decreases due to the absorption of electrons. The \textbf
{conversion efficiency} of each detector is defined as the ratio of the number of the photons that generate electrons after conversion
in any of the gas gaps to the number of the incident photons.

\subsection{Monte Carlo study for response of MRPC}
\label{}

The conversion electrons resulting from GEANT4 are used as the input to the Monte-Carlo based programme, developed to study
the response in the gas volume for signal generation. When the electron passes through the gas volume it ionizes the gas,
which leads to the avalanche formation. During primary ionization the average number of clusters formed into the gas mixture
is considered to be around nine/0.2 mm \cite{2}, which is the higher for slow electrons from converted photons than the value commonly used for minimum ionizing 
particles for the gas of given composition. The avalanche development from the primary cluster within the gas volume is 
governed by the Townsend coefficient ($\alpha$) and the attachment coefficient ($\eta$), which in this work are simulated by HEED 
\cite{8}. The probability to obtain \textit{n} avalanche electrons at position \textit{x} within the gas volume, considering the electrode to be at x=0, is given by the relation \cite{9}

\begin{eqnarray}
 P(n,x) &=& k \frac{\overline{n}(x) - 1}{\overline{n}(x) - k}~~,~~~~~~~~~~~~~~~~~~~~~~~~~~~~~~~~~n = 0\nonumber\\
&=& \overline{n}(x){(\frac{1 - k}{\overline{n}(x) - k})}^{2}{(\frac{\overline{n}(x) - 1}{\overline{n}(x) - k})}^{n-1}~,~~~~~n>0 
\end{eqnarray}
where $\overline{n}(x) = e^{(\alpha - \eta)x}$ and $k = \frac{\eta}{\alpha}$.

To reduce the computational time in avalanche formation by a large number of secondary electrons, in our case for $\overline{n}(x) > 200$, we have used the Central Limit Theorem (CLT), as mentioned in Ref \cite{9}, to obtain the average number of electrons and their spread after avalanche at a particular position. The space charge effect is also implemented by the application of a cut off on the number of avalanche electrons, after which multiplication stops. In the present case, the threshold applied is 1.6$\times$10$^{7}$. Finally the avalanche electrons induce current signals onto the MRPC pick-up strips. The particle is said to be detected if the induced charge crosses a threshold value, which is 20 fC in our case. The \textbf{detection efficiency} of the MRPC, in this work, is defined as the ratio of the number of particles crossing the threshold charge value to the number of converted  particles entering the gas gap. The concept of multi-gap has been implemented within the MC code by introducing a number of sub-gaps, keeping the total width of the gas volume same.

During this study, we have simulated the detection efficiency and the time response of MRPC for varying number of gaps. No effort has been made to implement the response of the readout electronics. We have defined the \textbf{avalanche growth time} as the time taken by the avalanche, starting from time t=0, to produce sufficient number of electrons within the gas volume so that it can induce a charge greater than 20 fC to the MRPC electrodes. The standard deviation of the distribution of the avalanche growth time taken over a large number of particles is called the \textbf{time resolution} (TR) of the detector for single photon. For a pair of detectors, facing each other, as shown in Fig. \ref{fig:1}., the difference of the time (t$_{1}$) taken by one photon to produce a detectable avalanche in one detector and the time (t$_{2}$) taken by the other photon to produce the same in another detector is  defined as the \textbf{pair time difference} [$\delta$t = t$_{1} \sim$ t$_{2}$]. The standard deviation of pair time difference taken over a large number of photon pairs is defined as \textbf{pair time resolution} (PTR) for pair of photons. For the case, when both the detectors are equally spaced from the point of origin of the photon pair, PTR is governed mainly by the folding of TR of two detectors.

\section{Simulation Results}
\label{}

\subsection{Efficiency of photon conversion}
\label{}
\begin{figure}[h!]
\begin{center}
 \includegraphics[scale=0.42]{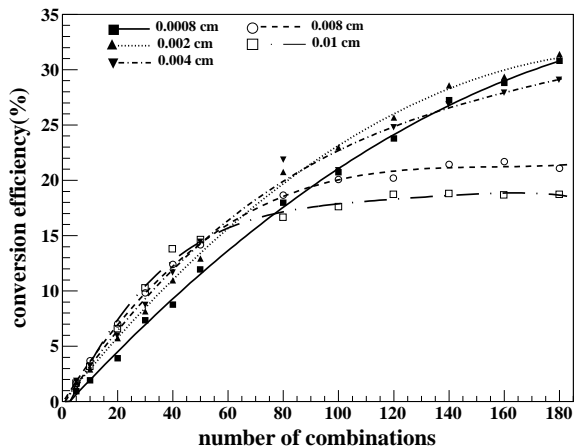}\\
\end{center}
\caption{\label{fig:epsart}Conversion efficiency as a function of number of gaps} 
\label{fig:2}
\end{figure}

The conversion efficiency is already defined in the previous section. Fig. \ref{fig:2}. shows the conversion efficiency over the number of gas
gap, where we have varied the converter thickness. 
By increasing the number of gaps, the effective thickness increases. It is seen clearly that with increasing number of gaps, the conversion 
efficiency increases. However for larger thickness, e.g. 0.1 mm, etc., the efficiency does not go higher than 18\% thereby suggesting the converted
electrons to stop inside the lead volume. Use of thinner converter in large numbers however increases the efficiency even further. The plot 
indicates that an efficiency >25\% can be achieved with a configuration having 120 gaps, the converter thickness being 0.002 cm. 
From the plot it can be seen that the converter having thickness 0.0008 cm is also giving an impressive conversion efficiency beyond 140 layers.
A saturation of the conversion efficiency for more than 140 gaps is observed for almost all the converter thickness studied. It is of utmost
importance to optimize the converter thickness for having a good conversion efficiency. During all through our study we have considered the 
optimized converter thickness as 0.004 cm.

\begin{figure}[h!]
\begin{center}
 \includegraphics[scale=0.42]{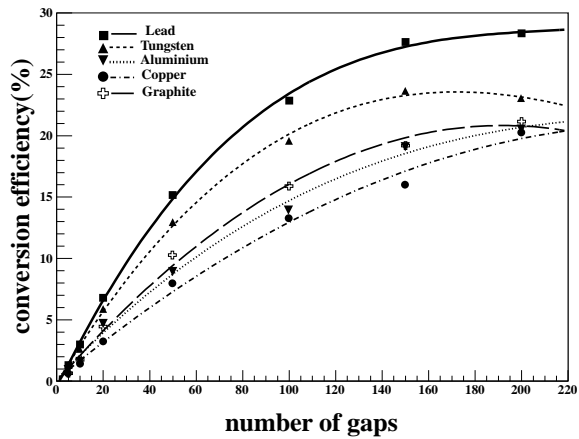}\\
\end{center}
\caption{\label{fig:epsart}Conversion efficiency as a function of number of gaps} 
\label{fig:3}
\end{figure}

Fig. \ref{fig:3}. indicates the photon conversion efficiency with the configuration prescribed earlier for different materials each having the 
same thickness of 0.004 cm. The plot clearly indicates that lead is the most suitable converter for 511 keV photons for 0.004 cm thickness of 
the converter layer.

We therefore conclude that a suitable tuning of converter material and thickness can help to achieve reasonable conversion efficiency in a multi-gap
configuration.

\subsection{Time response for MRPC}
\label{}

\begin{figure}[h!]
\begin{center}
 \includegraphics[scale=0.28]{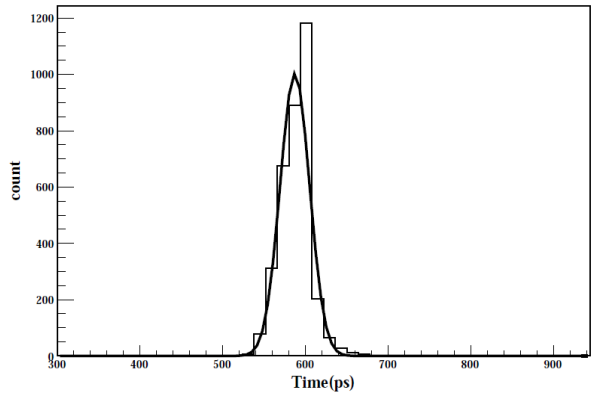}\\
\end{center}
\caption{\label{fig:epsart} Time resolution of the detector} 
\label{fig:4}
\end{figure}

In Fig. \ref{fig:4}. we have plotted the distribution of the signal collection time for single photon incident on the detector having 20 
gas gaps as obtained from MC simulation for MRPC response. The total number of entries in this plot indicate the total
number of the particles after conversion. The time resolution ($\sigma$) is found to be $\sim$ 17 ps , which is in good agreement with earlier 
results \cite{10}. 

\begin{figure}[h!]
\begin{center}
 \includegraphics[scale=0.40]{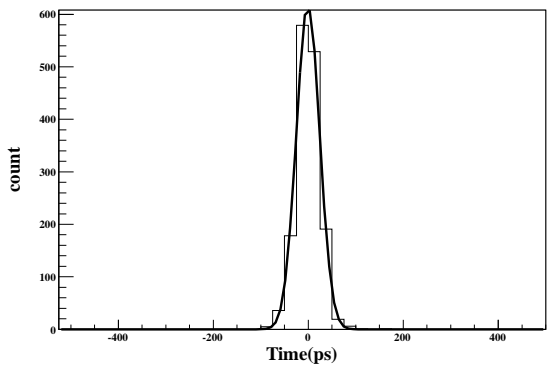}\\
\end{center}
\caption{\label{fig:epsart} Pair time resolution for both the detectors} 
\label{fig:5}
\end{figure}

The pair time resolution was calculated considering the start time to be zero. Fig. \ref{fig:5}. shows the distribution of the pair time difference
 collected from both the detectors. The standard deviation of the plot, 24 ps in this case, has a good agreement with the fact that both the 
detector have the same intrinsic time resolution.

We have observed the effect of the number of gaps in time resolution by varying the number of the gas gaps, keeping the total
width of the gas gap to be the same. It is seen that the time resolution decreases with the number of gas gaps.

\subsection{Estimation of the position resolution for an extended source of photon pairs}
\label{}

\begin{figure}[h!]
\begin{center}
 \includegraphics[scale=0.32]{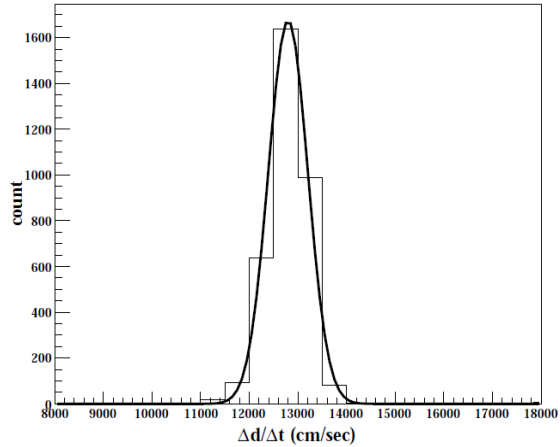}\\
\end{center}
\caption{\label{fig:epsart} Ratio of the diatance of the point of generation for the origin to the photon detection time by the system} 
\label{fig:6}
\end{figure}

In this work, we have also estimated the position resolution of the detector in the prescribed configuration for an extended annihilation 
source. For this purpose, we have introduced a Gaussian spread of the source position in one direction, e.g., in z-direction keeping mean at zero 
and sigma of 0.5 mm. The photon pairs are back to back on same plane. We have then obtained the corresponding time difference measured by the pair of 
detectors. In Fig. \ref{fig:6}. we have plotted the ratio of the distance of the source from the origin to the corresponding measured 
time difference. The spread gives an estimation of the resolution of the position measurement.

In this case the time of detection of a photon by the detectors is the sum of the time taken by the photon to reach the detector from the point
 of its origin and the signal generation time for each detector. It should be  mentioned that the contribution from the second plays the leading 
role in the spread of the detection time.

\section{Performance of the prototype MRPC}
\label{}

\begin{figure}[h!]
\begin{center}
 \includegraphics[scale=0.52]{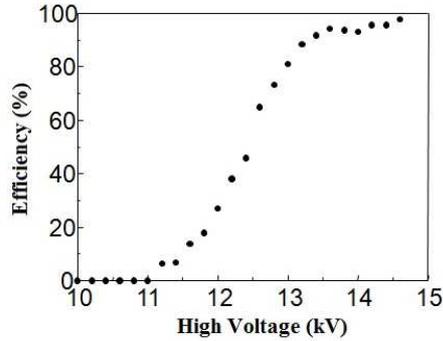}\\
\end{center}
\caption{\label{fig:epsart} Efficiency as a function of applied voltage} 
\label{fig:7}
\end{figure}

\begin{figure}[h!]
\begin{center}
 \includegraphics[scale=0.35]{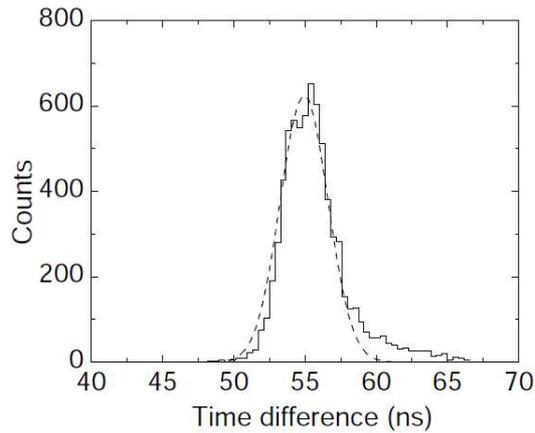}\\
\end{center}
\caption{\label{fig:epsart} Distribution of the time difference between the MRPC and the master trigger} 
\label{fig:8}
\end{figure}

The RPC technology mentioned in \cite{11,12} was used to build a four-gap Bakelite-based MRPC. Bakelite plates of 30 cm $\times$ 30 cm in dimension 
and thickness of 1.2 mm had been used as electrodes and a uniform spacing of 0.6 mm each had been maintained by a high resistive material G-10 
(resistivity $\sim$ $10^{13} \Omega$-cm). A mixture of Ar/i-butane/Freon(R134a) in the ratio of 55/7.5/37.5 was used as the working gas. 
A very thin coating of Silicone has been applied on the inner surfaces of the Bakelite plates to improve the surface roughness. The detector has 
been operated in streamer mode as it gives larger signal size and does not need any preamplifier. The detector has been tested with cosmic muons 
using the signals from three scintillators for trigger. Even though for these application, glass-based detectors operating in avalanche mode are
usually used, the performance of 4-gap Bakelite-based detector will provide the lower limit of resolution for a least expensive option of background
 filtering PET imaging. Fig. \ref{fig:7}. shows the efficiency as a function of the high voltage. An efficiency plateau > 95\% had been obtained for an applied high voltage 
greater than 13.5 kV. The distribution of the time difference between the master trigger and the signal from one MRPC strip is shown in Fig. 
\ref{fig:8}. From the time difference spectrum, the full width at half maximum (FWHM) and the corresponding standard deviation ($\sigma_{ij}$),
where i and j refer to scintillators and the MRPC, were obtained by fitting a gaussian function. $\sigma_{ij}$'s were similarly
obtained for 3 different pairs of the scintillators, taking START signal from one and STOP signal from the other. 
The intrinsic time resolutions of the MRPC and the scintillators were obtained from the individual standard deviations $\sigma_{i}$ ,
$\sigma_{j}$, which were extracted by solving the equation: ${\sigma_{ij}}^{2} = {\sigma_{i}}^{2} + {\sigma_{j}}^{2}$. 

\begin{figure}[h!]
\begin{center}
 \includegraphics[scale=0.55]{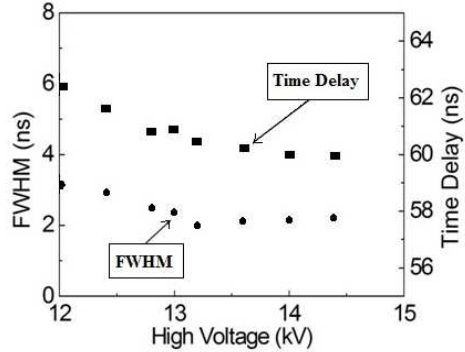}\\
\end{center}
\caption{\label{fig:epsart} Time resolution and time delay as a function of applied voltage} 
\label{fig:9}
\end{figure}
The variation of FWHM and the time delay for the MRPC is shown in Fig. \ref{fig:9}. The extracted time resolution ($\sigma$) of the MRPC 
at 14 kV operating voltage for a typical run was found to be  $\sim$ 870 $\pm$ 0.03 ns. 

\section{Conclusions and outlook}
\label{}
MRPC-based PET sysytem is potentially an attractive alternative to the scintillator-based sysytem. In this work we have performed detailed simulation
 for the detection of photon pairs for MRPC-based system. A layered structure of MRPC with thickness 
0.004 cm lead converter can give conversion efficiency greater than 25\%, which is higher compare to normally available values for gas detector systems. For a 
20 gap system, time resolution of photon pair can reach upto 17 ps. We have also performed simulation for the position measurement for 
an extended source of 500 $\mu$m width and found that the spread of the photon source can be measured with the addition of 3.17\% extra fluctuation. 
This suggests the appreciability of thse systems for measurements of photon-pair position to micron level.

As an effort to obtain a least expensive PET imaging system capable of filtering the background, we have built a 4-gap Bakelite-based MRPC giving time resolution of $\sim$ 900 ps. For applications requiring relatively lower resolution, such a system can be useful. This also 
provides a platform to build a full glass-based high resolution MRPC system for PET imaging.

\section{Acknowledgement}
\label{}

The authors gratefully acknowledge the technical contributions from Mr. G. Das and Mr. J. Kumar and also acknowledge Dr. S. Biswas, Dr. T. Ghosh and Dr. Sanjay Pal for their help in detector test as well as in the simulation.

\end{document}